\documentclass[trackchanges,twocolumn]{aastex701}

\newcommand{\Msun}{\mbox{M$_{\odot}$}}
\newcommand{\Lsun}{\mbox{L$_{\odot}$}}
\newcommand{\Rsun}{\mbox{R$_{\odot}$}}

\def\sun{\hbox{$\odot$}}
\def\earth{\hbox{$\oplus$}}

\def\ref{\hangindent=15pt     
        \hangafter=1}

\def\lesssim{\mathrel{\hbox{\rlap{\hbox{%
  \lower4pt\hbox{$\sim$}}}\hbox{$<$}}}}
\def\gtrsim{\mathrel{\hbox{\rlap{\hbox{%
  \lower4pt\hbox{$\sim$}}}\hbox{$>$}}}}

\usepackage{caption}
\usepackage{rotating}
\usepackage{amsmath}

\definecolor{jaredpurple}{RGB}{93, 63, 211}

\begin{document}

\title{Infrared spectroscopy of the longest-period Galactic Mira variables from Palomar Gattini-IR: Identification of a candidate thermally pulsing Super-Asymptotic Giant Branch star}

\author{Ming To Suen}
\affiliation{Department of Astronomy, Columbia University, 550 W 120th St. MC 5246, New York, NY 10027, USA}
\email{ms7106@columbia.edu}

\author{Kishalay De} 
\affiliation{Department of Astronomy, Columbia University, 550 W 120th St. MC 5246, New York, NY 10027, USA}
\affiliation{Center for Computational Astrophysics, Flatiron Institute, 162 5th Ave., New York, NY 10010, USA}
\email{kd3038@columbia.edu}

\author{Sam Rose}
\affiliation{Division of Physics, Mathematics, and Astronomy, California Institute of Technology, Pasadena, CA 91125, USA}
\email{}

\author{Rajorshi Bhattacharya}
\affiliation{Department of Astronomy, Columbia University, 550 W 120th St. MC 5246, New York, NY 10027, USA}
\email{}

\author{Viraj Karambelkar}
\affiliation{Department of Astronomy, Columbia University, 550 W 120th St. MC 5246, New York, NY 10027, USA}
\email{}

\author[]{Michael C. B. Ashley}
\affiliation{School of Physics, University of New South Wales, Sydney NSW 2052, Australia}
\email{}

\author[]{Cuc Dinh}
\affiliation{University of Hawai’i at Manoa - Institute for Astronomy, Honolulu, HI, USA}
\email{}

\author[]{Nicholas Earley}
\affiliation{Division of Physics, Mathematics, and Astronomy, California Institute of Technology, Pasadena, CA 91125, USA}
\email{}

\author{Danielle Frostig}
\affiliation{Center for Astrophysics $\vert$ Harvard \& Smithsonian, 60 Garden Street, Cambridge, MA 02138, USA}
\email{}

\author[]{Matthew Hankins}
\affiliation{Arkansas Tech University, 215 West O Street, Russellville, AR 72801, USA}
\email{}

\author[]{Charles~D.~Kilpatrick}
\affiliation{Center for Interdisciplinary Exploration and Research in Astrophysics (CIERA), Northwestern University, Evanston, IL 60201, USA}
\email{}

\author{Mansi M. Kasliwal}
\affiliation{Division of Physics, Mathematics, and Astronomy, California Institute of Technology, Pasadena, CA 91125, USA}
\email{}

\author[]{Anna M. Moore}
\affiliation{Research School of Astronomy and Astrophysics, Australian National University, Canberra, ACT 2611, Australia}
\email{}

\author{Roberto Soria}
\affiliation{College of Astronomy and Space Sciences, University of the Chinese Academy of Sciences, Beijing 100049, China}
\affiliation{INAF-Osservatorio Astrofisico di Torino, Strada Osservatorio 20, I-10025 Pino Torinese, Italy}
\affiliation{Sydney Institute for Astronomy, School of Physics A28, The University of Sydney, Sydney, NSW 2006, Australia}
\email{}

\author[]{Aswin Suresh}
\affiliation{Department of Physics and Astronomy, Northwestern University, Evanston, IL 60201, USA}
\affiliation{Center for Interdisciplinary Exploration and Research in Astrophysics (CIERA), Northwestern University, Evanston, IL 60201, USA}
\email{}

\author[]{Tony Travouillon}
\affiliation{Research School of Astronomy and Astrophysics, Australian National University, Canberra, ACT 2611, Australia}
\email{}

\begin{abstract}
The population of extremely long-period Mira variables in the Milky Way remains poorly characterized. Their rarity and heavy circumstellar obscuration have precluded systematic studies in optical surveys, yet they are expected to trace the most massive stars evolving through the Asymptotic Giant Branch (AGB). We present IRTF/SpeX spectroscopy of the longest-period ($P\gtrsim1000$\,d) Galactic Mira variables identified with the Palomar Gattini-IR survey. Twelve of thirteen targets exhibit characteristics consistent with massive O-rich AGB stars. In contrast, one peculiar source IRAS\,19552+3142 exhibits i) a large-amplitude ($\Delta J\gtrsim2$\,mag) pulsations with $P\approx1060$\,d, ii) strong C$_2$ and CO molecular bands indicating C/O$\gtrsim1$, and iii) SiC emission at $\approx11.3\,\mu$m in archival mid-infrared spectra. Its spectroscopically confirmed chemistry makes it the longest period C-rich Galactic Mira variable known; together with the rarity of similar objects, this source is difficult to reconcile with standard massive AGB evolution where hot-bottom burning maintains an O-rich surface composition. Instead, its properties may be explained by super-AGB evolution where dredge-out episodes produce a C-rich surface and circumstellar material. Its high luminosity ($M_{[4.5]}\lesssim-10$\,mag), red color ($[3.6] - [4.5]\gtrsim2.0$\,mag), and high mass-loss rate ($\gtrsim 10^{-4}\,\mathrm{M_\odot\,yr^{-1}}$) resemble extreme AGB star progenitors of intermediate-luminosity red transients in nearby galaxies and a known chemically peculiar super-AGB star candidate in the Small Magellanic Cloud. IRAS\,19552+3142 may thus provide a rare look at a Galactic super-AGB star in the thermally pulsing phase. Our results motivate improved theoretical predictions and long-baseline infrared searches for these short-lived, dust-enshrouded endpoints of stellar evolution.
\end{abstract}





\section{Introduction}

Asymptotic Giant Branch (AGB) stars represent the end fates of most stars in the universe \citep{Wood1999}. While lower mass ($\approx 1-2$\,\Msun) AGB stars are routinely identified as Mira variables with periods in the range of $100-300$\,d (e.g., \citealt{Sanders2023}), the evolution of higher mass AGB stars and their final remnants remains unclear \citep{Siess2006}. While the general evolution of AGB stars is well understood, the late evolutionary stages and final fates of the most massive AGB stars (initial masses $\gtrsim 7\,M_{\odot}$) remain poorly constrained \citep{Siess2006}. Very long-period Mira variables (periods $\gtrsim 1000$\,d; VLP Miras hereafter) are extremely rare, potentially because they represent a short-lived phase of stellar evolution \citep{Marigo2017} or because such long pulsation periods are attained only by the most massive AGB stars \citep{Feast2009, Groenewegen2020}.

As summarized in \citet{Menzies2019}, there are less than 20 known Galactic Mira variables in this period range, found very close to the Galactic plane -- supporting their association with the most massive AGB stars \citep{Whitelock2008} as predicted by theoretical models \citep{Trabucchi2019}. Many of these VLP Miras are also classified as OH/IR stars, which are dust-obscured, oxygen-rich AGB stars exhibiting circumstellar OH maser emission. OH/IR stars can exhibit particularly long pulsation periods and are often associated with higher-mass AGB populations \citep{jimenez2015study, nakagawa2022properties}. Extremely long period Miras have also been suggested to hide the long-sought populations of the most massive ($7 - 9$\,\Msun) `super' AGB stars (SAGB; \citealt{Siess2007, Doherty2014}) that ignite carbon under partially degenerate conditions, producing O–Ne–Mg cores and possibly electron-capture supernovae \citep{Thompson2009, Rose2025, Hiramatsu2021}.

Despite their significance, the identification of VLP Miras is difficult. Their location in the dusty Galactic plane, together with intense mass loss that enshrouds them in circumstellar dust, have precluded systematic identification in optical time domain surveys. No unambiguous SAGB star has been identified, largely because their luminosities and mass-loss properties overlap those of massive AGB stars (that have not undergone C-core ignition) and red supergiants \citep{Doherty2017, OGrady2020}.  Theoretical models predict that some SAGB stars may exhibit chemical peculiarities in their late evolution \citep{Doherty2014, OGrady2023} due to terminal dredge-up episodes, making them appear distinct from massive AGB stars. Therefore, systematic searches for VLP Miras in infrared (IR) time domain surveys together with spectroscopic characterization to probe chemical composition offer a powerful avenue to search for these elusive end-fates in stellar evolution.

The majority of the known VLP Mira variables are found in targeted surveys of the Large/Small Magellanic Clouds (LMC/SMC) and nearby galaxies. Early near-IR surveys of the LMC and SMC identified only a few dozen such stars \citep{Feast1989,Soszynski2009,Soszynski2011}. The longest period objects in these samples reach periods $P \approx 2000$\,d, and occupy the extreme bright and red end of the AGB sequence \citep{Groenewegen2009}. IR surveys such as Surveying the Agents of a Galaxy's Evolution (SAGE) and Dust in Nearby Galaxies with Spitzer (DUSTiNGS) expanded this population by uncovering numerous dust-enshrouded long-period Miras \citep{Riebel2010,Riebel2015,Boyer2011,Boyer2015}. The SPitzer InfraRed Intensive Transients Survey (SPIRITS) mid-IR monitoring program \citep{Kasliwal2016} revealed objects with periods extending to $\sim 3000$ days in nearby galaxies \citep{Karambelkar2019}; however, aliasing remains a limitation with period identification due to the sparse cadence. Due to their faintness at optical/near-IR bands, most sources remain without chemical characterization. Although some sources have probable chemical classifications with low resolution mid-IR spectroscopy or (primarily) photometry \citep{Sloan2016, Zijlstra2006, Groenewegen2018}, higher resolution confirmation is unavailable.

The census of VLP Miras within the Galaxy remains much less well charted, limited by the need for wide-area IR surveys of the Galactic plane. \citet{Soszynski2013} and \citet{Albarracin2025} presented catalogs of several thousand Galactic Miras towards the Galactic bulge and the far side of the disk from the Optical Gravitaional Lensing Experiment (OGLE; \citealt{Udalski2015}) and Vista Variables in the Via Lactea (VVV; \citealt{Minniti2010}) surveys respectively; however these catalogs did not include any objects with $P > 1000$\,d either due to different search criteria (e.g., focusing on Galactic kinematics with VVV data) or their red colors (which prevents discovery in the OGLE filters). \citet{Suh2021} and \citet{Groenewegen2022} presented all-sky catalogs of Galactic Mira variables using data from the Wide-field Infrared Survey Explorer (WISE) surveys \citep{Wright2010, Mainzer2014}, finding several dozen Galactic Mira variables with $P>1000$\,d. The very slow cadence of NEOWISE also makes identification of the fundamental pulsation periods difficult \citep{Suh2021, Suh2024}.

Although historically limited, the field of IR stellar time domain astronomy is opening up thanks to synoptic near-IR surveys. These surveys now offer pathways to a systematic census of the longest period objects for detailed chemical characterization. Recently, \citet{Suresh2024} presented the first systematic catalog of long-period variables identified from the Palomar Gattini-IR (PGIR) survey \citep{De2020a, Murakawa2024}. The catalog contains $>150,000$ Long Period Variables identified from $J$-band light curves, including several hundred with candidate periods $> 1000$\,d. Utilizing the same dataset, \citet{Earley2025} presented a catalog of the largest amplitude variable sources from the PGIR, which identified a number of the longest period Mira variables (owing to their large amplitude fundamental mode pulsations). The high cadence of PGIR ($\lesssim 3$\,d) allows us to conclusively identify fundamental mode pulsations, removing the severe contamination from both aliasing and long secondary period variations (see Appendix \ref{sec:app}; \citealt{Riebel2010, Riebel2015, Soszynski2009}). 

Here, we present near-IR spectroscopy of the thirteen VLP Miras confirmed from vetting of PGIR light curves of previously identified candidates -- to i) conclusively remove aliasing contaminants, ii) carry out chemical characterization and iii) compare against theoretical expectations for massive AGB evolution predicted by stellar models to understand their nature. Section \ref{sec:obs} describes the sample selection criteria and spectroscopic observations. In Section \ref{sec:analysis}, we discuss quantitative estimation of the pulsation periods and classification of their spectroscopic chemistry. Section \ref{sec:models} compares the derived chemical properties with models of stellar evolution, highlighting one source with peculiar chemistry and possible explanations. We end with a summary and future outlook in Section \ref{sec:summary}. 

\section{Observations and Data Reduction}
\label{sec:obs}

\subsection{Sample Selection}

\begin{figure*}[!ht] 
\centering 
\includegraphics[width=0.9\linewidth]{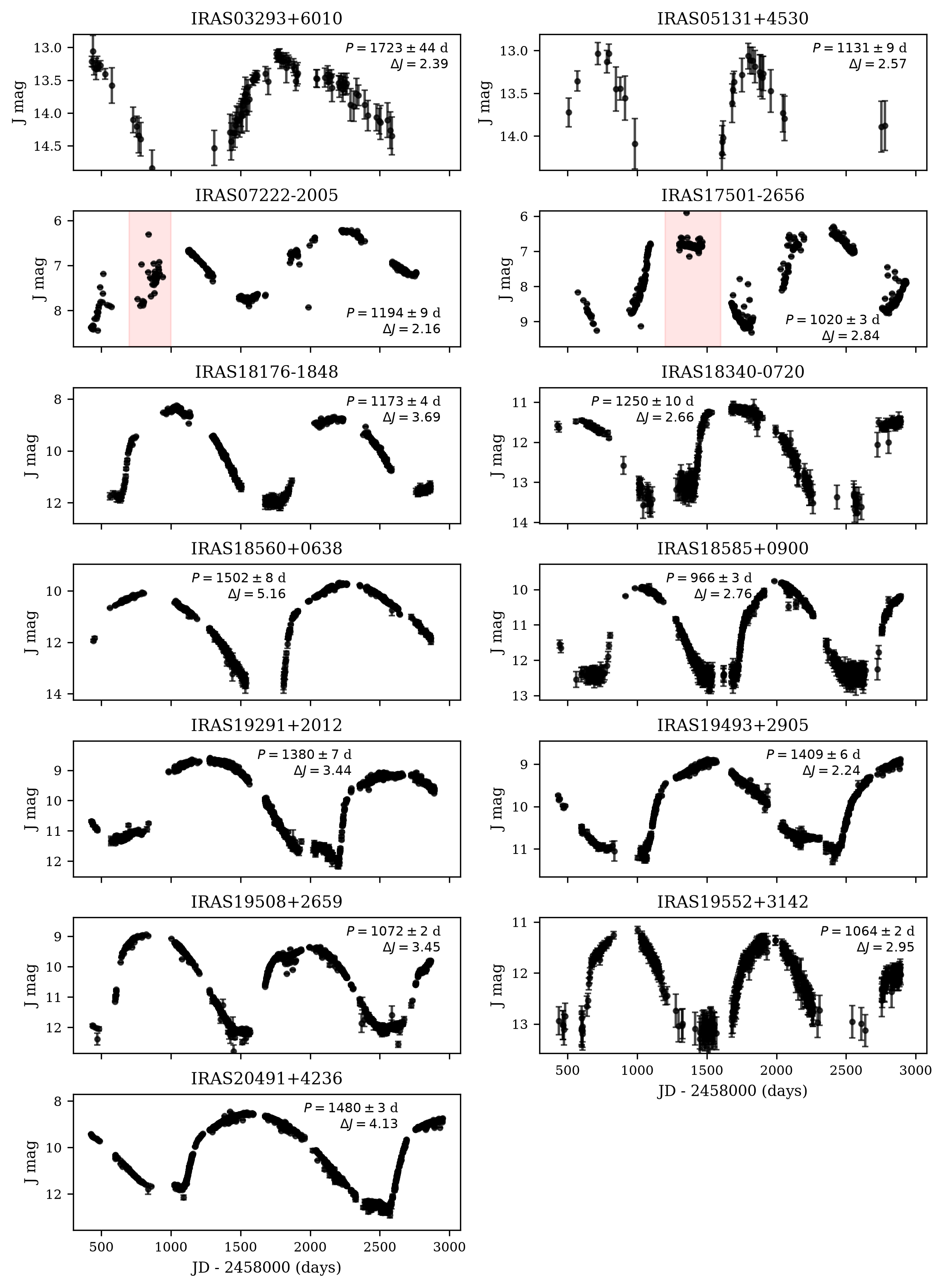} 
\caption{PGIR $J$-band light curves for the 13 VLP Mira variables in our sample -- black circles indicate observed photometry with error bars, together with the best-fit pulsation period ($P$) and amplitude ($\Delta J$). The red shaded regions include time periods where the source was nominally brighter than the survey saturation limit, and hence the photometry is unreliable. The saturation threshold was increased by $\approx 2$\,mag later in the survey due to modified detector readout \citep{De2020b}.} 
\label{fig:sample} 
\end{figure*}

Our targets were selected using data from the PGIR survey, which has monitored the entire visible sky since 2018 with a cadence of $\approx 1{-}3$ nights in $J$ band and a depth of $\approx 15$\,Vega\,mag. The sensitivity, cadence, and increasing baseline ($\gtrsim 2500$\,d) of PGIR make it especially well suited to identifying long-period, dust obscured, infrared-bright Mira variables that are otherwise invisible even in the deepest optical time domain surveys. To construct our sample of $P\gtrsim 1000$\,d Mira variables, we began with the recently published catalog of long-period variables \citep{Suresh2024}. We down-select this sample to those with reported \texttt{bestperiod} or \texttt{maxperiod} $\gtrsim 1000$\,d, which yielded 11187 candidates. Since this sample was created using only the first $\approx 1400$\,d of PGIR data, most of the reported sources were found to be spurious upon visual inspection of light curves. We therefore used the public light curve catalog \citep{Murakawa2024} together with more recent photometry that provides a total baseline of $\approx 2500$\,d.

Because very long period variability is also seen in semi-regular red giants, we visually inspected the light curves to identify sources with bona fide large-amplitude pulsations of i) amplitude $\Delta J \gtrsim 2$\,mag (removing red giants and long secondary period pulsations that have amplitudes $\lesssim 1$\,mag; \citealt{Soszynski2009}) and ii) $P\gtrsim1000$\,d, as approximately verified by the peak-to-peak separation (see Section \ref{sec:analysis} for subsequent quantitative analysis). This produced a sample of 5 sources. Because the catalog in \citet{Suresh2024} was created from earlier PGIR data, we also analyzed the recent compilation of large amplitude variables from \citet{Earley2025} to select longer period variables. This yielded an additional 2 sources with existing near-IR spectroscopic follow-up.

To complete our sample, we also examined the catalog of the 145 longest period variables with $P>1000$\,d identified from NEOWISE data by \citet{Groenewegen2022}. Because their sample was created from very slow cadence ($\approx 6$\,month) light curves, we extracted PGIR light curves of all the reported $P\gtrsim 1000$\,d sources in their sample within the PGIR footprint, identifying an additional 6 Mira variables that satisfy the same amplitude/period criteria. Details of the sample selection and contaminants are provided in Appendix \ref{sec:app}. Given that PGIR is limited by confusion in the most crowded Galactic plane fields \citep{De2020a}, we expect our sample to be complete for sources with mean brightness of $J\lesssim 12.5$\,Vega\,mag. The 13 light curves of the final sample\footnote{In addition to these targets, we also flagged long period pulsations and C-type chemistry in IRAS\,20003+3619; however it has been previously characterized as a DY-Per type H-deficient star \citep{Karambelkar2024, Earley2025} and not related to the evolved AGB stars discussed here.} are shown in Figure~\ref{fig:sample}. 

\subsection{Spectroscopic Observations}

\begin{figure*}[!th]
    \centering
    \includegraphics[width=1\linewidth]{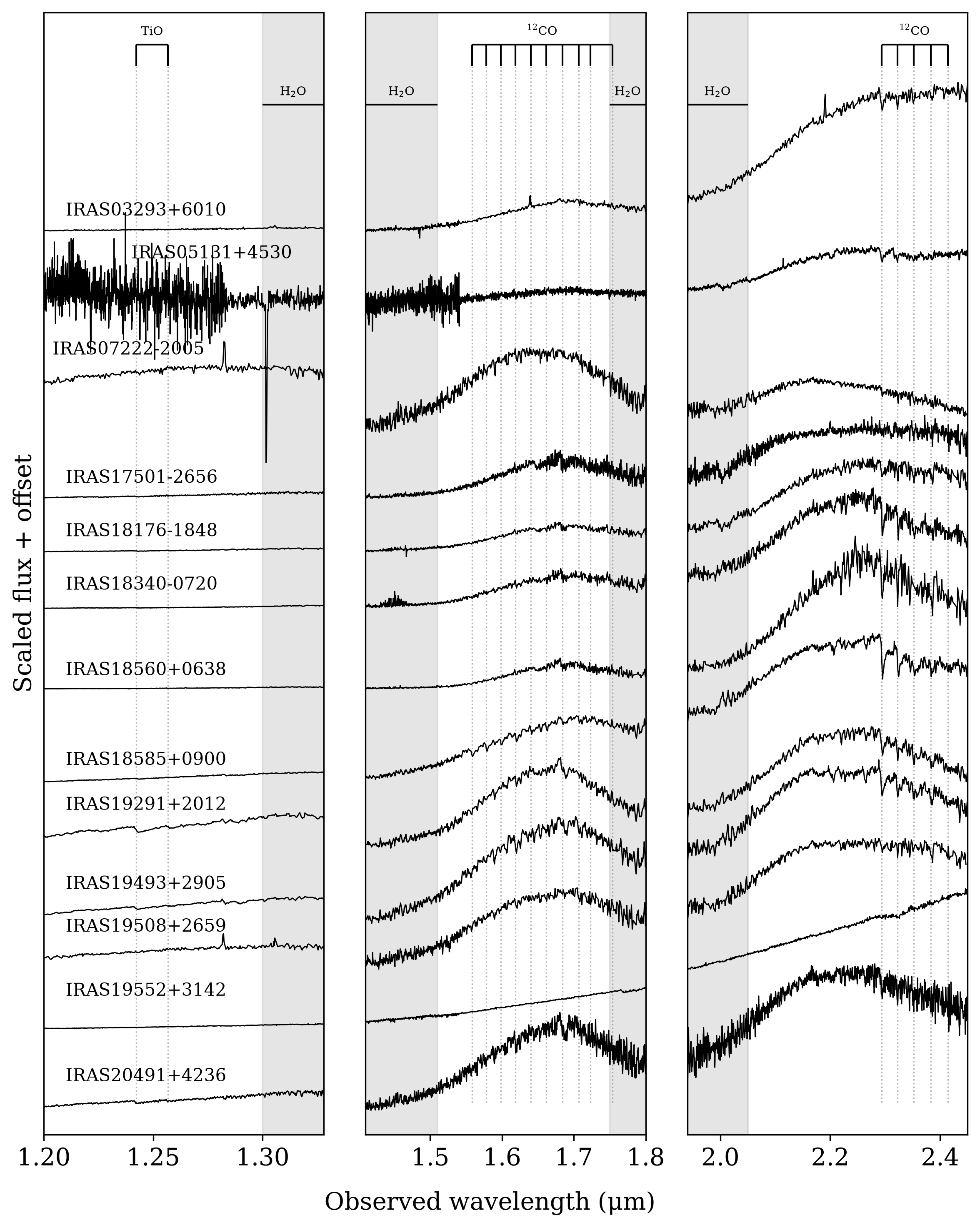}
    \caption{Near-IR spectra of the 13 targets in our sample (indicated by name), highlighting prominent molecular absorption features used for spectroscopic classification. The gray shaded regions indicate known regions of water absorption bands. }
    \label{fig:fullspec}
\end{figure*}

We carried out near-IR spectroscopic observations of the sources with the 3.0-m NASA Infrared Telescope Facility (IRTF) on Maunakea during semester 2025A (Program 2025A079; PI: De). We employed the SpeX spectrograph in the short cross-dispersed (SXD) mode, which provides a resolving power of $R \approx 2000$ with the $0.3''$ slit and simultaneous coverage of the $0.7{-}2.5\,\mu$m wavelength range \citep{Rayner2003}.  To achieve S/N $\gtrsim 50$ across the spectral range, we adopted AB nod sequences with individual exposures of 15 s. Typical sequences consisted of six exposures (ABABAB), giving a total integration time of $\approx 90$ s per target. Flats and arc frames were obtained every 2–4 targets, following standard calibration procedures. 

We reduced the data using the \texttt{pyspextool} package\footnote{\url{https://github.com/pyspextool/pyspextool}} \citep{pyspextool}, which performs flat-fielding, wavelength calibration, sky subtraction, and optimal spectral extraction. Wavelength calibration was achieved using internal arc lamps, while telluric correction was performed using observations of nearby A0V stars at similar airmass to each science target. The final spectra were flux calibrated using telluric standards. The new spectra were supplemented with previously published spectra of sources from the catalog of \citet{Earley2025}. Our final sample of spectra includes all 13 VLP Mira variables with $P\gtrsim 1000$\,d; their broadband spectra are shown in Figure \ref{fig:fullspec}.

\section{Analysis}
\label{sec:analysis}
\subsection{Period Estimation}

The well-sampled $J$-band light curves of the sources are shown in Figure~\ref{fig:sample}. Nearly all of these sources are not detected in contemporaneous optical surveys (such as the Zwicky Transient Facility; \citealt{Bellm2019}) due to their extremely red colors. The light curves predominantly show a primary, large amplitude ($\Delta J \gtrsim 2.0$\,mag) pulsation mode; in a few cases superimposed with higher frequency variability. We use a semi-parametric Gaussian process regression model including a fundamental mode pulsation, a higher frequency overtone and a long-term trend \citep{He2016} to derive the fundamental pulsation period. We use the \texttt{R} code distributed by \citet{He2016} to derive a primary pulsation period from the log-likelihood periodogram and a best-fit $J$-band light curve model including higher frequency variability terms. We estimate the statistical uncertainty in the period by simulating 10 realizations of the observed light curve by adding Gaussian noise scaled to the photometric uncertainties. 

\subsection{Spectroscopic classification}

\begin{figure*}[!ht]
    \centering
    \includegraphics[width=0.95\linewidth]{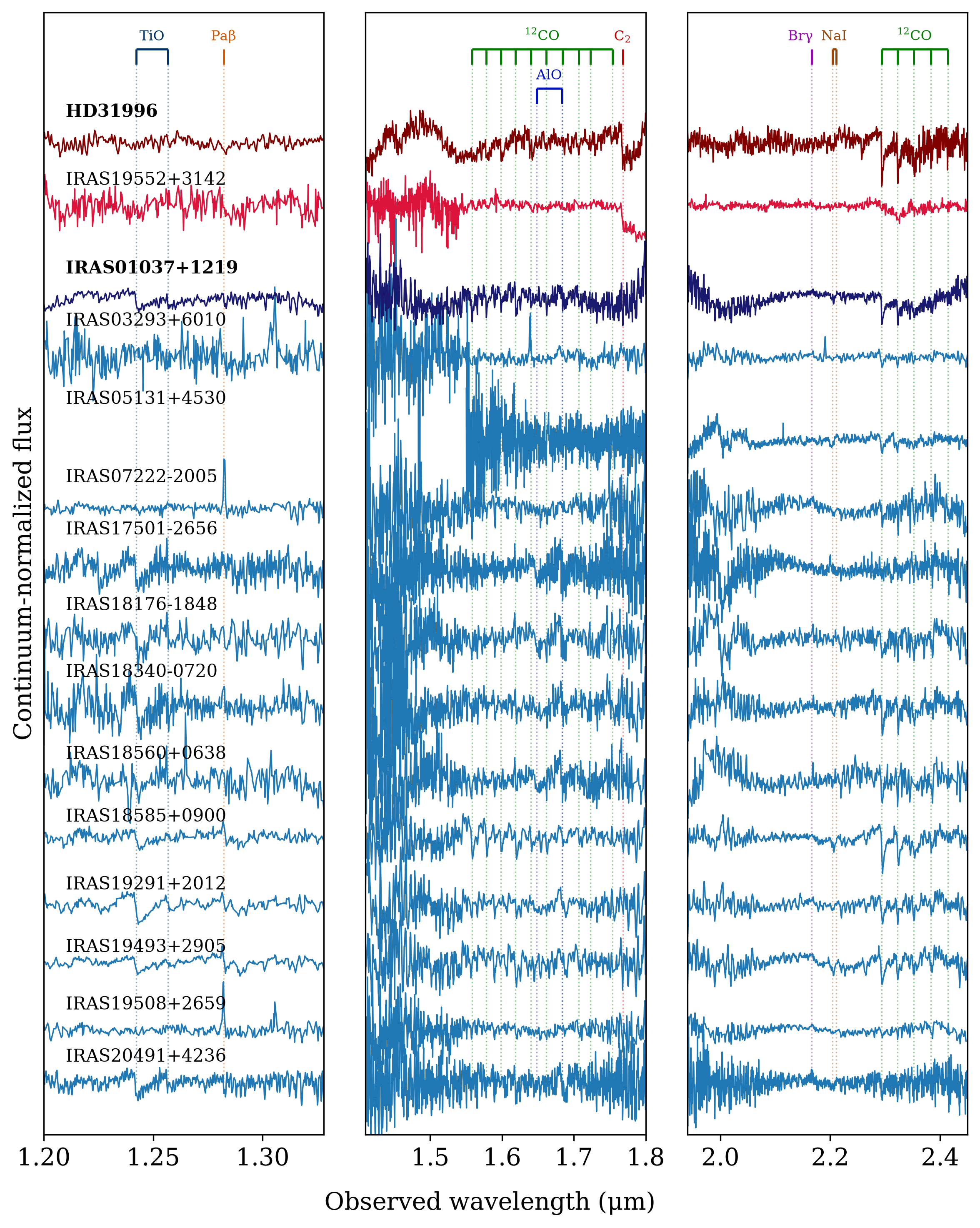}
    \caption{Continuum-normalized spectra of the sample of VLP Miras together with reference objects from the IRTF Spectral Library (HD\,31996 and IRAS\,01037+1219). Strong atomic and molecular absorption features are marked. For both the sample and reference objects, we indicate O-rich spectral types with blue/navy blue colors, and C-rich types with red/brown colors.}
    \label{fig:specnorm}
\end{figure*}

All of the sample sources exhibit red continua, commonly marked by inverted V-shaped broad absorption feature expected from water molecules. The only exception is IRAS\,19552+3142, which exhibits a steep red continuum with no observable water absorption.  To highlight the weaker features on top of the continuum, we show their continuum normalized spectra in Figure \ref{fig:specnorm}. We classified the 13 targets into O-rich or C-rich chemistry by identifying diagnostic molecular absorption features in the SpeX spectra using well-established spectroscopic indicators \citep{Rayner2009, Messineo2021} for evolved AGB stars. All sources except one (IRAS\,19552+3142) exhibit O-rich chemistry -- with dominant Titanium Oxide (TiO) bandheads in the $J$ band and deep water vapor (H$_2$O) absorption troughs at $1.4\,\mu$m and $1.9\,\mu$m (Figure \ref{fig:fullspec}). All sources exhibit deep CO first overtone absorption in $K$-band, characteristic of AGB spectra \citep{Rayner2009}.

The continuum normalized spectra exhibit more subtle differences between the spectra: three O-rich sources (IRAS\,17501-2656, IRAS\,18176-1848 and IRAS\,18560+0638) show clear absorption bandheads of AlO in the $H$-band, and one source (IRAS\,18585+0900) also exhibits noticeable $H$-band absorption band-heads of the CO second overtone. AlO is a rare molecule in astrophysical environments, but has been shown to be relatively common in AGB stars at very low temperatures and O-rich environments \citep{Banerjee2012} -- confirming the highly evolved nature of these sources. Two sources (IRAS\,07222-2005 and IRAS\,19508+2659) exhibit weak emission lines of H, e.g., Pa-$\beta$ near $1.28\,\mu$m and the Br series, commonly seen in pulsating Mira variables due to shocks in their extended atmospheres \citep{Tej2003, Jadlovsky2026}. IRAS\,19552+3142 is an outlier, exhibiting clear C-rich chemistry: it lacks continuum water absorption and other O-rich molecules; instead it only exhibits a clear C$_2$ absorption bandhead at $\approx 1.75\,\mu$m. Next, we compare the observed chemical properties to stellar evolution models in Section \ref{sec:models}, paying specific attention to the peculiar C-rich object. 

\section{Stellar evolution models}
\label{sec:models}

\subsection{Comparison to model grids}
We leverage existing stellar evolution models to compare the results of our chemical characterization against stellar theory. Past theoretical works make predictions for the evolution of periods of Mira variables as a function of initial mass and age -- with the surface chemistry being modified over time due to dredge-up episodes from the stellar interior. In general, stars with lower initial masses ($M_i \lesssim 2$\,\Msun) ascend the AGB without significant alteration of their surface chemistry \citep{ventura2018gas}. Therefore, they are predicted to evolve into relatively short period Miras ($P \lesssim 300$\,d) with surface chemistry close to solar composition ($C/O \approx 0.6$, producing O-rich chemistry). As more massive stars ($2$\,\Msun $\lesssim M_i \lesssim  4$\,\Msun) attain higher luminosities (and larger radii) in AGB evolution, strong mixing episodes due to the third dredge-up (TDU) are predicted to strongly enhance C-fraction at the surface producing C-rich Miras at periods reaching $\approx 500-600$\,d. For even more massive AGB stars, the effect of TDU is nullified due to the even higher core temperatures that are predicted to destroy the synthesized C via the CNO cycle (hot-bottom burning or HBB; \citealt{Herwig2005}), leading to the most massive AGB stars ($M_i \gtrsim 4$\,\Msun) attaining the longest periods ($\gtrsim 1000$\,d) with O-rich composition. 

\begin{figure}
    \centering
    \includegraphics[width=\linewidth]{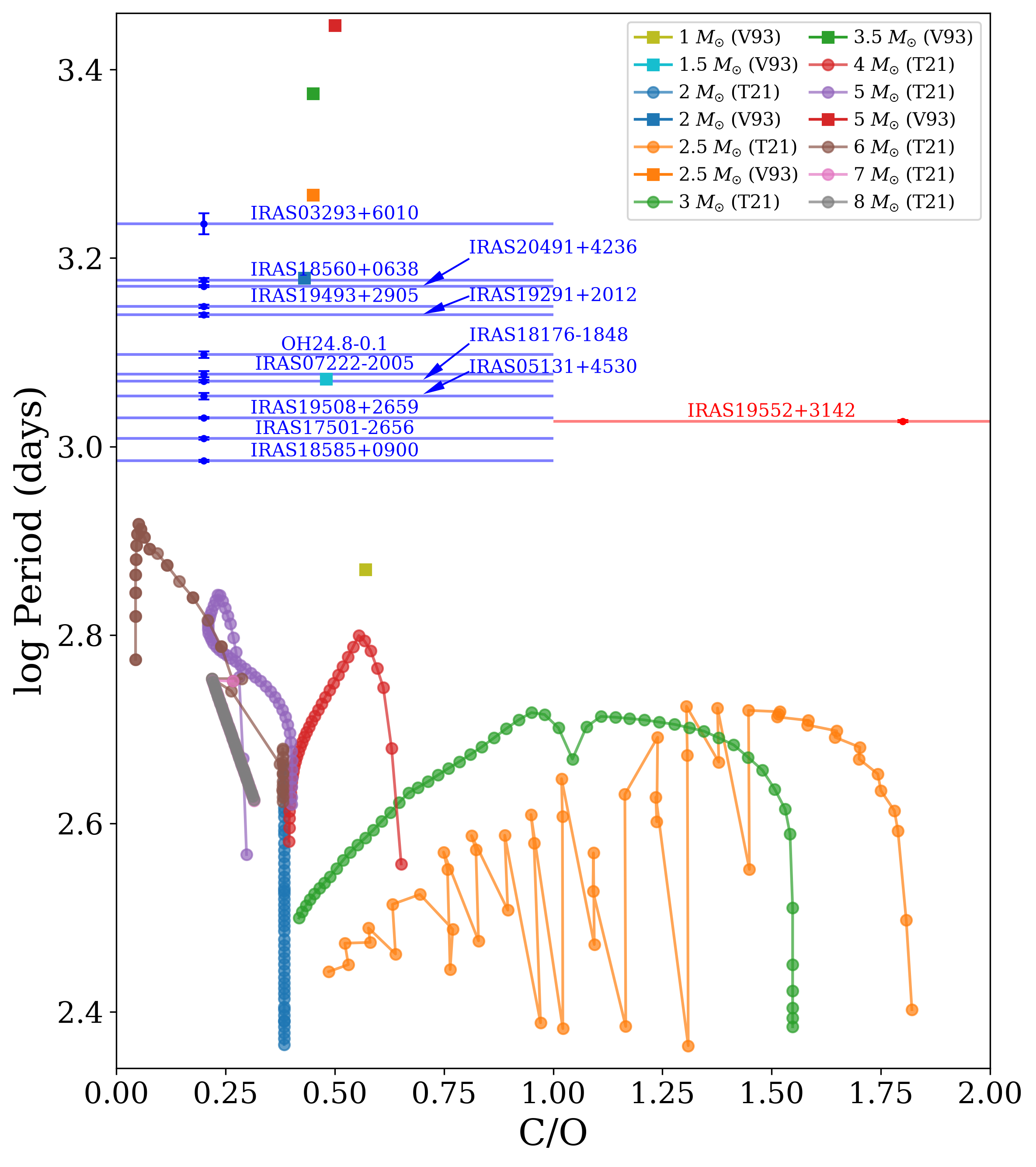}
    \caption{Theoretical predictions for the evolution of pulsation period with surface chemistry for Mira variables at different initial masses (see legend) from the models of \citet{Trabucchi2019, Trabucchi2021} (T21) and the terminal models of \citet{Vassiliadis1993} (V93). For comparison we overlay the pulsation periods of our sample with their inferred surface chemistry (C/O$< 1$ for O-rich and C/O$> 1$ for C-rich).}
    \label{fig:period_co}
\end{figure}

We summarize the expected evolution in Figure \ref{fig:period_co}, plotting the predicted Mira fundamental pulsation period as a function of surface chemistry. We generated isochrones using the CMD v3.9 interface\footnote{\url{https://stev.oapd.inaf.it/cgi-bin/cmd_3.9}}, utilizing PARSEC v1.2S tracks integrated with COLIBRI thermally pulsing AGB models \citep{Marigo2017}. The models incorporate standard circumstellar dust and mass loss prescriptions, with pulsation periods (fundamental mode, \texttt{pmode=0}) computed using the methods described in \citet{Trabucchi2019,Trabucchi2021}. In all cases, the stars evolve to increasingly longer pulsation periods during their evolution before mass loss terminates the AGB phase. The longest periods ($\approx 1000$\,d) are achieved for the most massive stars with $M_i \gtrsim 6$\,\Msun with O-rich surface composition. On the other hand, C-rich Miras arise in a mass range of $2-4$\,\Msun (for solar metallicity), reaching maximum periods of $\approx 600$\,d. For comparison, we also overlay the terminal pulsation period and C/O as a function of initial mass from the earlier solar metallicity models of \citet{Vassiliadis1993}. These models reach even longer periods of $\approx 2800$\,d with O-rich composition for massive AGB stars, although they do not reproduce C-rich chemistry at solar metallicity.

To compare our results to this suite, we also overlay the pulsation period and inferred surface composition for our VLP Miras; we employ approximate estimates for the C/O ratio based on the identified spectral features. As shown, the O-rich stars in our sample occupy the expected region for the longest period Mira variables expected from theoretical models. As such, these objects serve as a signpost for the most massive AGB stars currently in the Milky Way undergoing the TP-AGB phase. However, the peculiar C-type object IRAS\,19552+3142 is not explained by these models, with a pulsation period more than twice as long as the longest periods expected at C-rich chemistry. Its exceptionally long period may indicate a very large radius and, in turn, very high luminosity that is akin to the most massive AGB stars -- that has peculiarly turned C-rich in its terminal phase. In the next section, we further examine the C-rich candidate via modeling of its spectral energy distribution (SED) to estimate its physical properties. 

\subsection{Physical properties of IRAS\,19552+3142}

\begin{figure*}[!ht]
    \centering
    \includegraphics[width=0.49\linewidth]{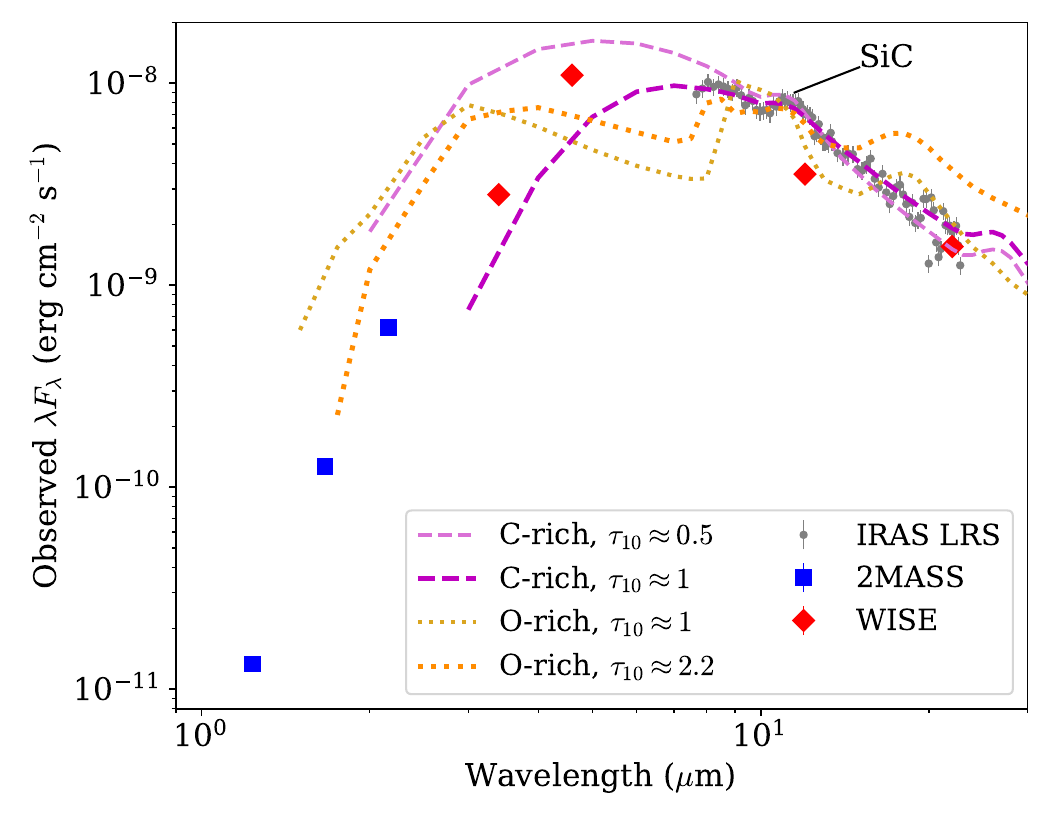}
    \includegraphics[width=0.49\linewidth]{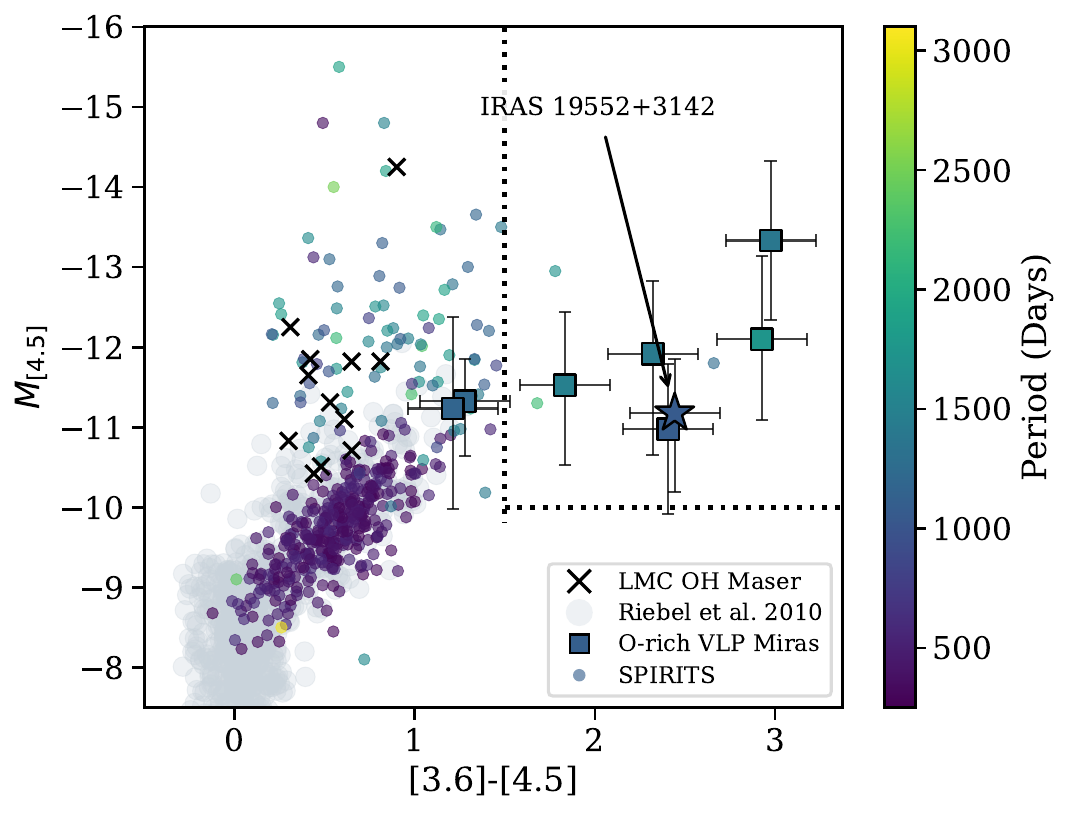}
    \caption{(Left) The observed SED of IRAS\,19552+3142 compared against radiative transfer models for dusty AGB stars of O-rich and C-rich chemistry \citep{Ventura2020}. We show two example optical depths ($\tau_{10}$ at $10\,\mu$m) for each case, highlighting the emission feature at $\approx 11.3\,\mu$m identified to be SiC. (Right) Mid-IR color magnitude diagram of Mira variables and OH masers in the LMC \citep{Riebel2010} compared to the extremely long period Mira variables from the SPIRITS survey \citep{Karambelkar2019}, the O-rich VLP Miras (squares) and IRAS\,19552+3142 (star) from this work. For the Galactic PGIR sources, we adopt the range of plausible distances indicated by Gaia astrometry and color uncertainty of $\approx 0.5$\,mag due to saturation in WISE data. The LPVs are color-coded by their pulsation period. The region separated by the dashed line indicates the space of extreme AGB star progenitors of SN\,2008S-like transients \citep{Thompson2009}. }
    \label{fig:seds}
\end{figure*}

We carry out a closer examination of the C-rich IRAS\,19552+3142\footnote{The source was previously identified as a C-star in \citet{Suh2024} based on WISE data but its pulsation period was incorrect.} to estimate its stellar and circumstellar properties. The source has prior near-IR photometry from the Two Micron All-sky Survey (2MASS; \citep{Skrutskie2006}), WISE \citep{Wright2010} as well as low resolution spectroscopy spanning $\approx 8-24\,\mu$m from the IRAS mission \citep{Kwok1997}. We compile these measurements and show their SED in Figure \ref{fig:seds}; we caution that the large variability of the source (Figure \ref{fig:sample}) makes it difficult to interpret SEDs compiled from multiple surveys. The source exhibits a steep red SED peaking at $\approx 5\,\mu$m, together with a clear emission feature in its IRAS spectrum near $\approx 11\,\mu$m\footnote{We do not correct the near-IR photometry for interstellar extinction since the total flux output is dominated by flux at $> 2\,\mu$m, given uncertainties with variability due to pulsation phase and interstellar dust models}. Both O-rich silicate molecules and C-rich SiC molecules have emission features in this region; \citet{Little-Marenin1987} previously associated this feature with SiC, suggesting a C-star classification \citep{Suh2024}. We carry out a closer comparison of the observed SED with radiative transfer models to constrain the circumstellar environment.

\citet{Ventura2020} presented a sample of IR SED models for C-rich and O-rich AGB stars of $\approx 1.5$\,\Msun and $\approx 5.0$\,\Msun. Although nominally constructed for lower mass AGB stars, the IR SED is sensitive primarily to the circumstellar chemistry and therefore their grid of SEDs as a function of optical depth (at $10\,\mu$m; $\tau_{10}$) is useful for comparison (as higher mass primarily increases the bolometric luminosity). Figure \ref{fig:seds} shows a comparison of the SED of IRAS\,19552+3142 to these models; both types of chemistry produce an observable feature around $\approx 11\,\mu$m. While the steep red color is matched by the higher $\tau_{10}$ model in the O-rich case, both O-rich models produce a prominent emission feature at $\approx 18\,\mu$m not observed in the data. The C-rich models at $\tau_{10} \approx 1$ reproduce the weak emission feature at $\approx 11\,\mu$m, the featureless slope between $12-20\,\mu$m as well as the red color, confirming a C-rich circumstellar dust envelope. 

We can translate the observed SED and inferred optical depth of the dust into physical parameters for the star. Integrating the observed SED of the source between $1-30\,\mu$m, we estimate a bolometric flux of $\approx 2.2\times 10^{-8}$\,erg\,cm$^{-2}$\,s$^{-1}$. The source is marginally detected in the optical bands, with a reported low significance parallax in Gaia DR3 of $0.21 \pm 0.19$\,mas (with RUWE $\approx 1.2$; \citealt{GaiaDR3}), corresponding to a distance range\footnote{Although Mira pulsation periods can be used to approximate distances, these are usually poorly calibrated or break down at very long periods \citep{Ita2011, Yuan2017}. We also note that a distance of $3.7 \pm 1.5\,\mathrm{kpc}$ has been reported for this source by \citet{bhattacharya2026mapping} using infrared photometry. However, since their distance estimation method was calibrated for O-rich LPVs, and adopting this distance would not affect our conclusions, and we retain the distance estimate from \citet{BailerJones2021}.} of $\approx 5.5^{+2.0}_{-2.1}$\,kpc (using the geometric distance estimate of \citealt{BailerJones2021}, since these extremely red objects have anomalous colors for photo-geometric estimates). Therefore, we estimate the $1-30\,\mu$m luminosity of the source to be $\approx (1 - 4) \times 10^4$\,\Lsun. Assuming a stellar temperature of $\approx 2500$\,K expected for extremely cool AGB stars \citep{Suh2025}, the corresponding stellar radius is $R_* \approx 500-1000$\,\Rsun. The inferred optical depth ($\tau_{10} \approx 1$) can be related to the mass loss rate ($\dot{M}$) of the star as 
\begin{equation}
\begin{split}
\dot{M} \approx{} & 10^{-4}\,M_\odot\,{\rm yr^{-1}}
\left(\tau_{10}\right)
\left(\frac{550\,{\rm cm^2\,g^{-1}}}{\kappa_{10}}\right) \\
& \times \left(\frac{R_{\rm in}}{1.5\times10^{14}\,{\rm cm}}\right)
\left(\frac{v_d}{15\,{\rm km\,s^{-1}}}\right)
\left(\frac{10^{-3}}{r_{dg}}\right)
\end{split}
\end{equation}
where $\kappa_{10}$ is the dust opacity (per unit dust mass; \citealt{Zubko1996}) at $10\,\mu$m, $R_{in}$ is the inner radius of the dust shell (assumed to be at $\approx  2- 5 \times R_*$), $v_d$ is the wind velocity and $r_{dg}$ is the dust to gas mass ratio\footnote{We note that uncertainties in the dust to gas ratio may introduce order unity corrections in the inferred mass loss rate}.

\subsection{The nature of IRAS\,19552+3142}

Motivated by the combination of the long pulsation period ($\approx 1060$\,d) and peculiar C-rich composition of IRAS\,19552+3142, our modeling reveals the following characteristics: i) the C-rich photospheric composition is accompanied by an extended C-rich dust shell, ii) a high luminosity in the range $\approx (1-4) \times 10^4$\,\Lsun and iii) an elevated mass loss rate $\sim 10^{-4}$\,\Msun\,yr$^{-1}$. For comparison, distance estimates for a subset of the O-rich VLP Miras from \citet{bhattacharya2024distance} place them in the relatively nearby Galactic disk, with luminosities broadly comparable to that of IRAS\,19552+3142 (see also Figure \ref{fig:seds}). Despite these similarities, their contrasting chemistry remain difficult to reconcile with the models discussed above, in which stars reaching the longest periods remain O-rich as HBB prevents C dredge-up.

We attempt to understand its nature by drawing comparisons to both previously observed objects and theoretical models. IRAS\,19552+3142 shares some similarities to the class of objects known as the luminous carbon stars in the Magellanic Clouds \citep{vanLoon1999}. Their proposed explanation is that mass loss reduces the envelope mass until HBB ceases, allowing further TDU episodes to make the star carbon-rich before it leaves the AGB. The brightest object in that study, IRAS\,04496-6958, has a C-rich atmosphere but retains an O-rich dust component, consistent with the cessation of an earlier O-rich wind. IRAS\,19552+3142 instead shows carbon-rich chemistry in both its atmosphere and the dust traced by its mid-IR spectrum -- indicating a relatively long-lived ($\gtrsim 10^4$\,yr) HBB turn-off phase. However, should HBB turn-off be universal in the late-phase evolution of the most luminous (and longest period) VLP Miras, we would expect to see a higher fraction of C-rich VLP Miras; instead, 12 of our 13 VLP Miras are O-rich, suggesting that any late HBB turn-off phase has to be restricted to a subset of these stars. Further theoretical work is therefore needed to confirm the duration of the HBB turn-off and if it can produce luminous stars with both C-rich photosphere and circumstellar dust. 

An alternative explanation is that the long period reflects envelope expansion rather than a particularly high initial mass. \citet{DellAgli2021} suggested that common-envelope (CE) interaction with an engulfed companion could explain the large radii and high mass-loss rates of extremely red carbon stars in the Magellanic Clouds \citep{Gruendl2008}. In this picture, the rise in molecular opacity after C/O exceeds unity may expand the star, potentially initiating interaction with a companion, and explaining both the large radius and strong mass loss inferred for IRAS\,19552+3142. This model is likely inconsistent for two reasons.

First, Figure \ref{fig:seds} compares our sources with LMC Mira variables and OH masers \citep{Riebel2010}, and the long-period variables from the SPIRITS survey \citep{Karambelkar2019}. We also show the region occupied by the extreme AGB star progenitors of intermediate luminosity red transients, as defined by \citet{Thompson2009}. Most of our sources fall within this region given their distance and color uncertainties, despite having different chemistries. Thus, these extreme colors and luminosities do not require a transition to C-rich chemistry, and therefore binary interaction driven by C-opacity expansion is not a prerequisite to producing very long period red objects. Second, the longest-period C-rich Galactic\footnote{\citet{Groenewegen2018} provide a list of 5 C-rich Mira variables in the LMC with $P > 1000$\,d; however some of these may be unreliable since some periods come from low-cadence NEOWISE data (e.g., Appendix \ref{sec:app}) and uncertain color-based chemical classification for very red objects \citep{Menzies2019}.} Mira variable previously reported in the literature has a pulsation period\footnote{The Galactic source V0829\,Cas has been suggested as a C-Mira with a period of 1060\,d \citep{Groenewegen1998} but its classification remains unclear due to lack of available details \citep{Feast2006}.} of 725\,d \citep{LeBertre1992}. IRAS\,19552+3142 would therefore represent the longest-period (by $> 50$\%) C-rich AGB star identified in the Milky Way. The lack of known stars in this period gap\footnote{Since Mira variables become particularly dust enshrouded only at the longest periods, we suggest that objects with $P< 1000$\,d are likely more complete in past optical characterization efforts} is problematic for the CE interpretation, since we would expect it to be populated by systems spanning a range of binary separations and periods. Therefore, IRAS\,19552+3142 possibly represents a distinct, rare evolutionary chain for forming C-stars at extremely long periods.

Finally, we suggest the possibility that IRAS\,19552+3142 is an SAGB star that has ignited carbon in its core. Because very massive AGB stars are expected to undergo HBB that prevents dredge-up of C in the TP-AGB phase, SAGB evolution offers an alternative pathway to produce C-rich surface chemistry in prior evolution. Before the thermally pulsing phase, some massive SAGB models undergo dredge-out: a helium-burning convective region meets the hydrogen-rich convective envelope \citep{Jones2016}, thereby mixing freshly produced carbon to the surface \citep{Doherty2017}. While subsequent HBB destroys carbon, the star remains C-rich due to the enhancement prior to the TP-AGB phase. This mechanism has been suggested to be a possible explanation for some C-rich extremely metal poor stars with enhancement of s-process elements \citep{Jones2016} as well as possible pre-SAGB stars in the Magellanic Clouds \citep{OGrady2020, OGrady2023}.

 A close analog of IRAS\,19552+3142 may be the extreme AGB star MSX SMC 055 (IRAS\,00483-7347) in the SMC (see also \citealt{Kastner2008} for possible candidate in the LMC). This source was previously classified as a large amplitude C-rich star \citep{Kraemer2009} based on optical and Spitzer/IRS spectra, and subsequently shown to be a very long-period, luminous Mira variable ($P \approx 1800$\,d; \citealt{Groenewegen2009, Groenewegen2018}) with directly measured enhancement of Rb \citep{GH2009}, indicating activation of the s-process and a possible SAGB nature. While this channel may naturally explain the C-rich atmosphere and circumstellar dust in IRAS\,19552+3142\footnote{The extremely red SED of IRAS\,19552+3142 makes it very challenging to obtain a Rb measurement at 7800\,\AA.}, we note that this source is likely less luminous but more dust-obscured than MSX SMC 055, indicating a lower mass or earlier phase of evolution. The universality of this phase and its dependence on mass and metallicity therefore require further theoretical investigation \citep{Jones2013, Doherty2014}; as the only Galactic analog of MSX SMC 055, future detailed studies of IRAS\,19552+3142 would be crucial to understand its circumstellar environment (e.g. via radio maser lines).

The possible SAGB origin is also supported by the progenitors of intermediate-luminosity red transients, which have been proposed to be electron-capture supernovae from SAGB progenitors (also known as SN\,2008S-like transients; \citealt{Hiramatsu2021,Thompson2009}), and the progenitors of some nearby Type-II supernovae from low-mass supergiant progenitors. Late-time JWST spectroscopy of AT\,2019abn revealed Class C PAH features indicating carbonaceous circumstellar material \citep{Rose2025}. Similarly, \citet{Kilpatrick2025} reported the detection of graphitic dust\footnote{See also \citet{vanDyk2026} who claimed O-rich dust around the progenitor of SN\,2025pht.} around the progenitor of the Type II SN\,2025pht -- indicating a massive star ($\approx 15$\,\Msun) surrounded by C-rich circumstellar medium. IRAS\,19552+3142 shares this broad chemical property and lies in the color-magnitude region associated with SN\,2008S-like progenitors (Figure \ref{fig:seds}), although its colors are substantially bluer than the progenitor of SN\,2025pht. Its high inferred mass-loss rate is also consistent with a heavily obscured progenitor -- suggesting the presence of a common mechanism that likely substantially enhances the terminal mass loss rate for all these sources.

\section{Summary and Future Prospects}
\label{sec:summary}

We combined long-baseline PGIR photometry with near-IR spectroscopy from IRTF to identify and characterize Galactic very long period Mira variables that have historically remained out of the reach of optical time domain surveys. Our main results are:
\begin{itemize}
    \item We confirmed 13 large-amplitude Mira variables with pulsation periods $P\gtrsim1000$\,d. The dense cadence and $\approx2500$\,d baseline of PGIR allow us to distinguish genuine long-period pulsations from aliases and spurious periods in earlier catalogs.

    \item Twelve of the thirteen sources exhibit O-rich spectra, with prominent H$_2$O and TiO absorption. Their long periods and surface chemistry are broadly consistent with massive AGB stars undergoing hot-bottom burning. Several also show AlO absorption or hydrogen emission associated with their cool, pulsating atmospheres.

    \item IRAS\,19552+3142 is the sole C-rich source, with a pulsation period of $\approx1060$\,d, making it the longest-period C-rich AGB star identified in the Milky Way to date. It exhibits photospheric C$_2$ absorption, while its mid-IR spectrum independently supports carbon-rich circumstellar dust, including SiC emission near $11.3\,\mu$m. This combination is not reproduced by the standard AGB model grids available in the literature.

    \item The SED and plausible distance range of IRAS\,19552+3142 imply a luminosity of $\approx(1-4)\times10^4$\,\Lsun, a radius of $\approx500-1000$\,\Rsun, and a mass-loss rate of order $10^{-4}$\,\Msun\,yr$^{-1}$ under the adopted wind and dust assumptions. These estimates remain limited by the uncertain distance, but nonetheless confirm a heavily enshrouded, luminous AGB star with C-rich chemistry.

    \item We discuss possible explanations for IRAS\,19552+3142 -- including a late carbon-rich phase following the cessation of HBB and envelope expansion enhanced by binary interaction. Neither scenario readily explains why most stars in our sample are O-rich or why VLP C-rich Miras are so rare. These mechanisms must therefore operate in only a small fraction of stars. 
    
    \item Instead, we discuss the possibility of a SAGB progenitor involving dredge-out before the TP-AGB phase, resulting in C-rich surface chemistry even if HBB prevents further enrichment of C. Its chemistry and circumstellar properties make it an interesting analog of dust-obscured transient progenitors. Improved luminosity and abundance measurements, together with predictions for the duration of these phases, are needed to distinguish these possibilities.
\end{itemize}

A more complete census requires cadenced IR time-domain surveys spanning multiple pulsation cycles, coupled with spectroscopy to establish surface and circumstellar chemistry. Many additional VLP Miras likely remain undetected toward the Galactic bulge, where source confusion limits PGIR sensitivity and therefore a census of their total numbers. Current surveys such as the PRime-focus Infrared Microlensing Experiment (PRIME; \citealt{Sumi2025}) will be particularly well suited to finding and monitoring this population over a long baseline in a high metallicity environment. Similarly, the upcoming Cryoscope project will complete the census of these bright IR variables in the southern Milky Way \citep{Kasliwal2025}. Although similarly luminous Galactic sources will generally saturate surveys with the {\it Roman Space Telescope}, their extragalactic counterparts offer a promising route to measuring the incidence of these rare phases and clarifying the final evolution of massive AGB and SAGB stars. In particular, {\it Roman} surveys of nearby galaxies will extend these searches to populations with well-constrained distances, complemented by JWST mid-IR spectroscopy to establish the photospheric and circumstellar chemistry. 


\begin{acknowledgments}
We are grateful to John Rayner (IRTF Director) for supporting the service mode observations used for this program. We thank Lynne Hillenbrand for reviewing an earlier version of this manuscript and providing feedback. We acknowledge the support of the National Aeronautics and Space Administration for WISE data analysis through ADAP grant number 80NSSC24K0663. Palomar Gattini-IR (PGIR) is generously funded by Caltech, Australian National University, the Mt Cuba Foundation, the Heising Simons Foundation, the Binational Science Foundation. PGIR is a collaborative project among Caltech, Australian National University, University of New South Wales, Columbia University and the Weizmann Institute of Science. We acknowledge Astronomer observing with the Infrared Telescope Facility, which is operated by the University of Hawaii under contract 80HQTR24DA010 with the National Aeronautics and Space Administration.
C.D.K. and A.S. gratefully acknowledge support from the NSF through AST-2432037, the HST Guest Observer Program through HST-SNAP-17070 and HST-GO-17706, and from JWST Archival Research through JWST-AR-6241 and JWST-AR-5441.
\end{acknowledgments}

%
\facilities{WISE, PGIR, IRTF, IRAS}



\bibliography{sample701}{}
\bibliographystyle{aasjournalv7}



\appendix

\section{Sample selection and contaminants}
\label{sec:app}

\citet{Groenewegen2022} provides a recent summary of the sample of VLP Mira variables known in the Milky Way. The previous largest list was cataloged in \citet{Menzies2019}, who reported a sample of 17 Galactic sources with $P > 1000$\,d. Nearly all of these sources do not have previously reported near-IR spectroscopy; however, we focus our work on objects where the periods can be confirmed with PGIR data. 16 of these sources lie within the PGIR observing footprint, while only 4 are sufficiently bright in $J$-band for reliable PGIR light curves in crowded Galactic fields. We verified the periods of these objects, and included these objects in our sample (2 of these were already included based on the compilation from \citealt{Suresh2024} and \citealt{Earley2025}). 

\begin{figure}[!h]
    \centering
    \includegraphics[width=\linewidth]{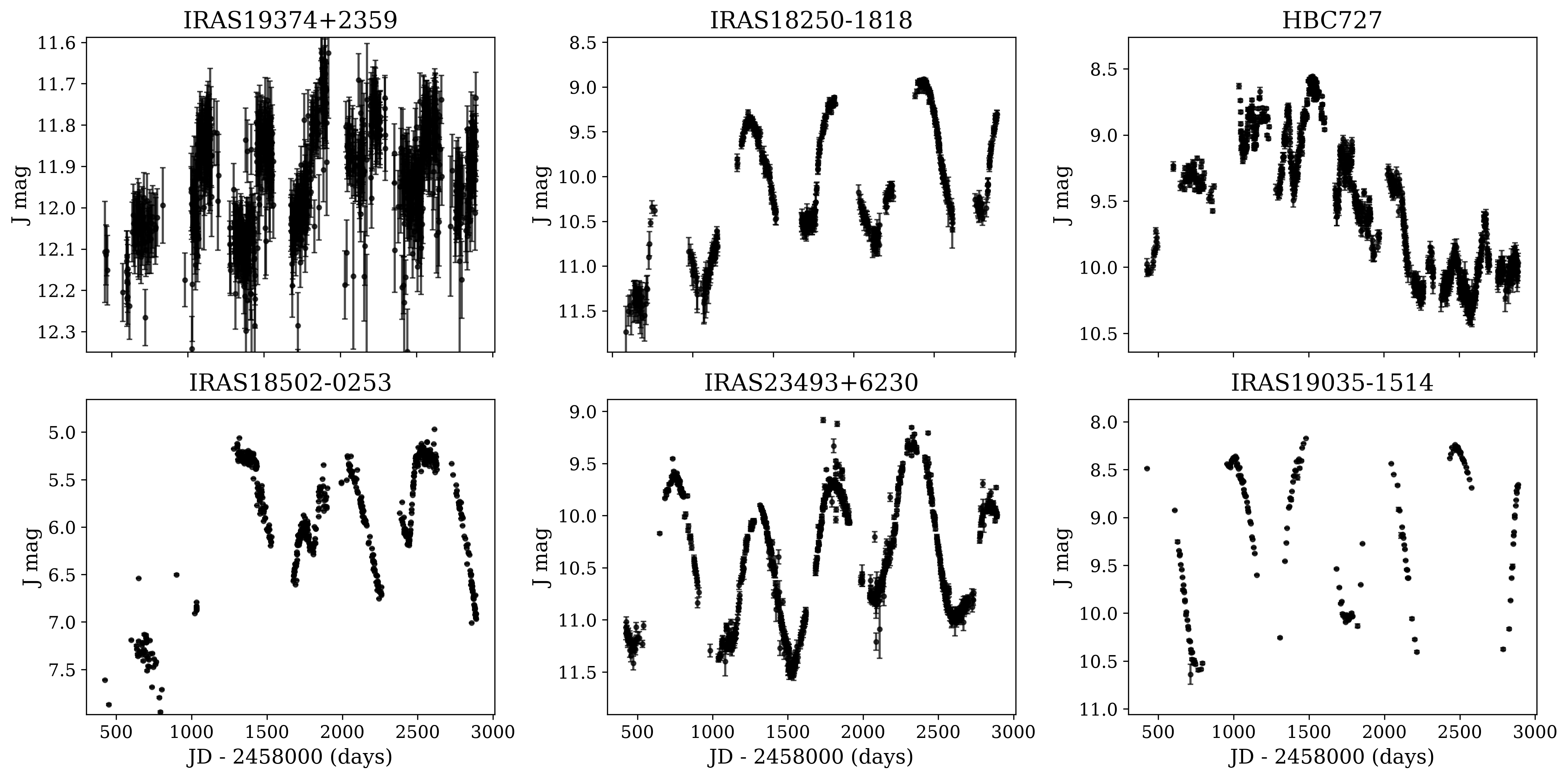}
    \caption{Examples of PGIR $J$-band light curves of sources that were suggested to have $P > 1000$\,d from NEOWISE data, but clearly have shorter periods based on the higher cadence ground-based coverage.}
    \label{fig:contam}
\end{figure}

\citet{Groenewegen2022} presented a sample of 154 VLP Mira candidates based on NEOWISE time series. However, the slow cadence of the NEOWISE survey and the severe saturation for these bright mid-IR sources requires re-examination of the claimed periodicities with PGIR light curves (where possible). We generated $J$-band light curves of 78 sources that fall within the PGIR observing footprint -- while most of these sources are too faint in $J$-band for photometric measurements, we find significant contamination of this sample with objects that have clearly shorter periods. Figure \ref{fig:contam} shows examples where the ground-based, higher cadence data clearly indicate shorter period variability. Of the 19 sources with reliable PGIR time series, 9 have $P > 1000$\,d; 3 of these 9 sources were already included in our sample based on the \citet{Menzies2019} and \citet{Suresh2024} catalog, resulting in the addition of 6 new sources. We also investigated the reported three C-Miras with $P > 1000$\,d (identified from WISE light curves) in \citet{Suh2024}: one of the three sources (IRAS 16328-4656) lies outside the PGIR footprint, one is too faint in near-IR (the unconfirmed V0829\,Cas reported in \citealt{Groenewegen1998}) while one object (RAFGL 996) clearly shows shorter timescale pulsations in its PGIR light curves reported in \citet{Earley2025}.

\end{document}